\documentclass[twocolumn,prl,showpacs]{revtex4}

\newcommand{\nn}{\nonumber}
\renewcommand{\H}{{\cal H}}
\renewcommand{\S}{{\vec{S}}}
\renewcommand{\L}{{L^\prime}}
\newcommand{\K}{{\cal K}}
\newcommand{\vac}{|\Omega\rangle}
\newcommand{\bra}{|\Omega\rangle}
\newcommand{\ket}{\langle\Omega|}
\renewcommand{\xi}{{J}}
\newcommand{\xish}{\xi^{{\rm HSh}}}
\newcommand{\Stot}{{\cal S}}

\begin{document}

\preprint{YerPhI-1577(2)-2002}

\title{ On the ground state properties of antiferromagnetic half-integer \\
 spin chains with long-range interactions }

\author{Tigran Hakobyan}

\email{hakob@lx2.yerphi.am}

\affiliation{
   Theory Division, Yerevan Physics Institute,
   375036, Alikhanyan 2, Yerevan, Armenia
 }

\date{\today}

\begin{abstract}

The Lieb-Shultz-Mattis theorem is extended to Heisenberg chains
with long-range interactions. We prove that the half-integer spin
chain has no gap, if it possesses {\it unique} ground state and
the exchange decays faster than the {\it inverse-square} of
distance between spins. The results can be extended to a wide class
of one-dimensional models.

\end{abstract}


\maketitle

The zero-temperature ground state properties of one-dimensional
antiferromagnets have been studied intensively over a long period.
There are various numerical and field theoretical approaches to
the problem. At the same time there are just few exact
non-perturbative methods. The spin $S$=1/2 antiferromagnetic (AF)
Hei\-sen\-berg spin chain with nearest-neighbor (n.n.)
interactions was solved exactly by Bethe \cite{b31}. Bethe's
ansatz was applied later to peculiar higher spin chains with
polynomial spin exchange \cite{b83}.

Lieb, Schultz and Mattis \cite{lsm61} offered another
non-perturbative approach. They proved that $S$=1/2 AF n.n.
Heisenberg chain has {\it gapless} excitations above the {\it
unique} ground state. Later Haldane \cite{h83}, using
$\sigma$-model description, argued, that the AF n.n. {\it integer}
spin Heisenberg chains have a {\it gap} between the ground state
and the first excited state, whereas the {\it half-integer} spin
chains have {\it no gap}. Haldane's conjecture was proven exactly
for the {\it half-integer} spin chains \cite{k85,al86} by means of
extension of the Lieb-Schultz-Mattis (LSM) theorem. The LSM
theorem was extended also to the case with an applied magnetic
field \cite{oya97_1}. It was generalized to Heisenberg chains with
$SU(n)$ \cite{al86} and other \cite{l01} symmetries, to fermionic
chains \cite{oya97_2}, to spin \cite{fsk97} and fermionic
\cite{ghr98} models on ladders, etc. Very recently, its extension
to higher dimensions has been discussed \cite{o00_1,o00_2,ml00}.

This letter is devoted to the extension of LSM theorem to AF
Heisenberg chains with {\it long-range} interactions. The $S=1/2$
chain with the spin exchange decaying as $1/r^2$, where $r$ is the
distance between spins, has been extensively investigated since it
was solved exactly by Haldane and Shastry \cite{hsh88}. The
integrability is a consequence of an infinite symmetry of this
model \cite{hhtbp92}. Despite of presence of long-range
interactions, its low-energy properties are similar to those of
the chain with n.n. interactions: both models are gapless and
belong to the same universality class. Here we discuss the
zero-temperature ground state properties of the {\it
translationally} invariant AF {\it half-integer} spin chain with
{\it arbitrary} exchange coupling. We will prove that, in the
thermodynamic limit, it either has {\it gapless} excitations or
{\it degenerate} ground states, provided that the coupling decays
faster than $1/r^2$.

The Hamiltonian of one-dimensional Heisenberg chain with
long-range interactions is
\begin{equation}
\label{HL} \H_L = \sum_{r=1}^{\L}\sum_{i=0}^{L-1}\xi(r)\;
\S_i\cdot \S_{i+r}, \quad \L= (L-1)/2,
\end{equation}
where the chain length $L$ is chosen to be {\it even}.   Note,
that the spin exchange coupling $\xi(r)$ depends on the distance
between the spins only. Thus, $\H_L$ is translationally invariant.
The periodic boundary conditions are assumed and the sum over $r$
is taken up to the half-size $\L$ of the chain.

Let us consider the nonlocal unitary operator
\begin{eqnarray}
\label{U} U=e^{{\cal A}},   \qquad  {\cal A}=\frac{2\pi
i}{L}\sum_{k=1}^{L}k S^z_k .
\end{eqnarray}
It rotates the spins around the $z$-axis with the relative angle
$2\pi/L$ between the neighboring spins. We are interested in
formal algebraic properties of $U$. However, as it was shown
recently, it has a clear physical meaning too. In spin chains and
ladders, the ground state expectation value of $U$ can be treated
as an order parameter, which characterizes various
valence-bond-solid ground states \cite{nt02}. In sine-Gordon
theory, it is  related to expectation values of vertex operators
\cite{nv02}.

 It is well known that the
action of (\ref{U}) on the ground state of the  Heisenberg chains
with {\it short-range} interactions gives rise to the state with
an energy, which approaches the energy of the ground state in the
thermodynamic limit $L\to\infty$ \cite{lsm61,al86}. Below we will
discuss the possibility to extend this property to the Hamiltonian
 (\ref{HL}) with {\it long-range} interactions.

In this letter, for the sake of simplicity, we assume the {\it
uniqueness} of the ground state $\vac$ of finite chain. Note, that
in this case $\vac$ is a spin-{\it singlet}. Our assumption is
fulfilled for wide class of models on even length chains. In case
of alternating couplings, i.e. $\xi(2r-1)>0,\ \xi(2r)\le0$, it can
be proven {\it exactly} using Perron-Frobenius type arguments
\cite{al86}.

It is convenient to rewrite $\H_L$ in the following form:
\[
\H_L=\sum_{r=1}^{\L}\xi(r)\H_L^r, \quad {\rm where} \quad
\H_L^r=\sum_{i=0}^{L-1}\S_i\cdot \S_{i+r} .
\]
Then the energy gap between the ground state $\vac$ and the
shifted state $U\vac$ is
\begin{eqnarray}
\label{DeltaEL} \Delta E_L &=& \sum_{r=1}^{\L} \Delta E_L^r,
\qquad
{\rm where} \\
\Delta E_L^r &=& \ket U^+\H_L^rU\bra - \ket\H^r_L\bra . \nn
\end{eqnarray}
The straightforward calculations show that
\begin{eqnarray}
 \label{UHlU}
   U^+\H_L^rU - \H^r_L &=&   \sum_{i=0}^{L-1}  \left\{  \K(r/L)
    S_i^+ S_{i+r}^- + {\, \sf h.\; c.} \right\}  \nn   \\
&+& \, [\H_L^r,{\cal A}], \qquad  {\rm where}  \\
  \K\left(\phi\right)&=&\frac12\{\exp(2\pi i\phi) - (1+2\pi
i\phi)\}
  \nn
\end{eqnarray}
\noindent and $S_i^\pm=S_i^x\pm iS_i^y$. Note, that the
coefficients in front of spin operators don't depend on the
absolute positions of the spins, but depend on their distance.
Taking the sum of (\ref{UHlU}) over $r$, we obtain:
\begin{eqnarray}
\label{UHU}
   &&  U^+\H_LU - \H_L =  \sum_{r=1}^{\L}\xi(r)\left\{ \K(r/L) {\H_L^r}^+
     + {\sf h. \; c.} \right\} \nn \\
   &&  \quad     + \, [\H_L,{\cal A}], \quad {\rm where}\quad
     {\H_L^r}^\pm =  \sum_{i=0}^{L-1} S_i^\pm S_{i+r}^\mp.
\end{eqnarray}
 The mean value of the commutator $[\H_L,{\cal A}]$
vanishes on $\vac$, because $\vac$ is an eigenstate of $\H_L$. The
mean value of ${\H_L^r}^{\pm}$ is of the order of $L$:
\begin{equation}
\label{HpmUpper}
  |\ket{\H_L^r}^\pm\bra| \le 1/4 L .
\end{equation}
Now, using (\ref{DeltaEL}), (\ref{UHU}) and (\ref{HpmUpper}), we
can estimate the upper bound of the gap:
\begin{eqnarray}
\label{Delta} |\Delta E_L| &\le&
\frac{L}2\sum_{r=1}^{\L}\xi(r)|\K(r/L)| \nn\\
&=& \frac{1}{L}\sum_{r=1}^{\L}r^2\xi(r)\frac{|\K(r/L)|}{2(r/L)^2}.
\end{eqnarray}
 The function $|\K(\phi)|/(2\phi^2)$ is continuous on the unit
interval $0\le\phi\le1$ (see the definition in (\ref{UHlU})). So,
it can be replaced in (\ref{Delta}) by its maximal value $C$ and
we obtain:
\begin{equation}
\label{D} |\Delta E_L| \le \frac{C}{L}\sum_{r=1}^{\L}r^2\xi(r).
\end{equation}
We conclude, that the energy gap between the ground state and the
shifted state $U\vac$ vanishes in the thermodynamic limit,
provided that
\[
\lim_{L\to\infty}\frac{1}{L}\sum_{r=1}^{\L}r^2\xi(r)=0. \] Any
function, obeying
\begin{equation}
\label{xil}
 \lim_{L\to\infty}r^2\xi(r)=0,
\end{equation}
satisfies this condition. Note, that up to now we haven't
distinguished between the integer and the half-integer spin
chains.

In order to prove that $U\vac$ is a real excitation, i.e. it
doesn't coincide with the ground state $\vac$ and doesn't approach
it in the thermodynamic limit, it is sufficient to show that both
states are orthogonal \cite{lsm61,al86}. We use the translation
$T$ by one lattice site to demonstrate this. The ground state is
an eigenstate of $T$, because $T$ commutes with the Hamiltonian
(\ref{HL}): $T\vac=e^{ i p}\vac$, where $ p=0,\pi$ due to the
reflection symmetry. The shift operator (\ref{U}) transforms as:
$T^{-1}UT =U\exp(2\pi i S_1^z) \exp\left( -2\pi
      i \Stot^z/{L}
     \right)$,
where $\Stot^z=\sum_{i=0}^{L-1}S_i^z$ is the $z$-component of the
total spin $\vec{\Stot}$. We have $\Stot^z\vac=0$, because $\vac$
is a singlet.
 Using the equations above, we obtain:
$T U \bra = e^{-2\pi i S}e^{ i p} U \bra$. We see, that the
eigenvalues of $T$ on $\vac$ and $U\vac$ differ for {\it
half-integer} $S$ only. So, in this case the two states are
orthogonal.  Then, in the $L\to\infty$ limit, either {\sf i)}
$U\vac$ generates another ground state(s) or {\sf ii)} $\vac$
remains unique ground state with gapless excitations, created by
$U\vac$. In the second case, of course, the ground state should be
a spin-{\it singlet}.

We come to the conclusion that:

{\it For the translationally invariant AF Heisenberg half-integer
spin chain with long-range interactions the low-energy state,
approaching the ground state in the thermodynamic limit, exists,
if the exchange coupling of interacting spins decays faster than
the inverse-square of their distance. If, in addition, the ground
state remains unique, gapless excitations appear above it. }

This statement excludes the possibility to have a gap and unique
ground state simultaneously under the aforementioned conditions.
This fact remains true even if we omit the ground state uniqueness
requirement for finite $L$. The examples of $\xi(r)$, which
satisfy the conditions above, are:
\begin{eqnarray*}
&& {\sf a)}\ 1/(r^2\log(r)); \qquad {\sf b)}\ 1/r^\alpha, \,
\alpha>2; \\
&& {\sf c)} \ r^n\exp(-r/a),\,a>0.
\end{eqnarray*}
The last case corresponds to an effective short-range interaction.

The statement above gives only the {\it sufficient} condition for
the chain to be gapless or to have degenerate ground states in the
thermodynamic limit. The integer spin chains or chains with the
exchange coupling decay slower than $1/r^2$ can have gap and
unique ground state simultaneously.  They can also be gapless. For
example, the $S=1$ chain with the alternating exchange
$\xi(r)=(-1)^r/r^\alpha, \ 1<\alpha<3$ exhibits gapless behavior
\cite{sv97}. However, the same chain with $\xi(r)=1/r^2$ has a
gap, as it was shown numerically in \cite{sv96}. Its ground state
properties are similar to those of Haldane chain with n.n.
interactions \cite{h83}.

As we have mentioned before, the $S=1/2$ chain with $\xi(r)=1/r^2$
corresponds to the Haldane-Shastry integrable model. In case of
the {\it periodic} boundary conditions the coupling is slightly
modified \cite{note_L}: $ \xish_L(r)=\left((L/\pi)\sin(\pi
r/L)\right)^{-2}$. For $\xi(r)=\xish_L(r)$ the upper bound of the
gap in (\ref{Delta}) is nonzero. This fact doesn't guarantee the
existence of a low-energy state \cite{note}. Nevertheless, the
exact solution shows that this model has gapless excitations.

Very recently, the LSM theorem is considered in higher dimensions
\cite{o00_1,o00_2,ml00}, where the gap estimation method, used in
this letter, can't be applied. Nevertheless, the reasonable
arguments, used by Laughlin \cite{l81} in quantum Hall effect, can
be presented in support of LSM theorem in this case too. These
arguments are based on the assumption that during the certain
adiabatic process the gap remains robust. This process consists in
the insertion and the subsequent increment of an fictitious
magnetic flux, which penetrates the ring of the fermionic chain.
In the spin language, the magnetic flux is equivalent to a twist
of the boundary conditions. It would be interesting to clarify, if
the same argument can be applied in the presence of long-range
interactions. This would be a way to extend LSM theorem to the
chains with the exchange couplings, which decay slower than
$1/r^2$.

The results in this letter can be generalized to a wide class of
one-dimensional chains with long-range interactions, including
fermionic chains, spin chains with higher symmetries, with an applied
external field, etc.

The author is grateful to Prof. R.~Flume for hospitality during
his visit at Physikalisches Institut, Universit\"at Bonn. He
thanks R.~Flume, R.~Poghossian and D.~Karakhanyan for stimulating
discussions.

 This work is supported by Volkswagen Foundation of
Germany, grants INTAS \#00-561, INTAS \#99-01459 and Swiss SCOPE
grant.


\begin{thebibliography}{99}
\bibitem{note_L}
Of course, we have: $\lim_{L\to\infty}\xish_L(r)=1/r^2$. Note,
that the dependence on the chain length doesn't change anything in
the final result (\ref{Delta}), where we can impose
$\xi(r)=\xi_L(r)$.
\bibitem{note} For $\xi(r)=\xish_L(r)$ the right hand side of (\ref{Delta}) is equal
to the nonzero finite value of the integral
\[
\frac{\pi^2}2\int_0^{1/2} \frac{|\K(\phi)|}{\sin^2(\pi
\phi)}d\phi.
\]
\end{thebibliography}
\end{document}